\documentclass[showpacs,preprintnumbers,nofootinbib,letterpaper,11pt]{revtex4}

\usepackage{amsmath}

\usepackage{dcolumn}
\usepackage{epsfig}
\usepackage{verbatim}
\usepackage{psfrag}
\usepackage{bm}
\usepackage{bbm}
\usepackage{latexsym}
\usepackage[english]{babel}
\usepackage{amsthm}
\usepackage{color}
\usepackage{amssymb}

\begin{document}

\preprint{ITP-UU-11/15, SPIN-11/09}

\title{Negative Energy Cosmology and the Cosmological Constant}

\author{Tomislav Prokopec}
\affiliation{Institute for Theoretical Physics (ITP) \& Spinoza
Institute, Utrecht University, Postbus 80195, 3508 TD Utrecht, The
Netherlands \\ \texttt{\textup{T.Prokopec@uu.nl}}}

\begin{abstract}
It is well known that string theories naturally compactify on anti-de Sitter spaces,
and yet cosmological observations show no evidence of a negative 
cosmological constant in the early Universe's evolution.  
In this letter we present two simple nonlocal modifications of the 
standard Friedmann cosmology that can lead to observationally viable cosmologies with an initial (negative) cosmological constant. The nonlocal operators we include are toy models for the quantum cosmological backreaction. In Model I an
initial quasiperiodic oscillatory epoch is followed by inflation and a late time matter era, representing a dark matter candidate. The backreaction in Model II quickly compensates the negative cosmological term 
and the Universe ends up in a late time radiation era.
\end{abstract}


%

\maketitle

\section{Introduction}
\label{Introduction}

 Inflationary cosmology~\cite{Guth:1980zm}
is currently the leading paradigm that describes the very
early Universe. Inflation is driven by a substance that can be well approximated by a positive cosmological constant, and in practice it is usually realised by the potential energy of a (nearly) homogeneous scalar field, dubbed the inflaton. In fact, it is not known what is the precise nature of the substance that
drives inflation, nor it is known what was the Universe alike before inflation, close to the Planck scale at which (nonperturbative) quantum gravity effects become important. String theory as the leading contender for quantum gravity is of little help here, since the desire to preserve supersymmetry after compactification to four space-time dimensions
favors the compactification on six dimensional Calabi-Yau spaces on anti-de Sitter (AdS)  backgrounds, {\it i.e.} on space-times with a negative cosmological constant $\Lambda$. This is clearly at odds with inflation, which requires (a substance that can be well mimicked by) a positive cosmological constant $\Lambda$.~\footnote{We do not consider fixes that construct metastable string vacua~\cite{Kachru:2003aw} as satisfactory, since this type of models require a large amount of fine tunning.} In this letter we construct a nonlocal cosmology that resolves this impasse. We show that in our model strong backreaction effects can lift the Universe from an anti-de Sitter phase to an inflating Universe. We also show that the subsequent inflationary era can last long enough (i.e. more than 60 e-foldings) to solve all of the standard problems inflation is invoked to solve, which include the homogeneity, isotropy and causality problems, as well as that it can provide nearly scale invariant seeds for the large scale structure of the Universe. Finally, we show that the late time cosmology is that of matter era in Model I and of radiation era in Model II. Within our toy models we do not have a simple explanation for the dark energy of the Universe.

\section{Standard Cosmology}
\label{Standard Cosmology}

 Let us now for completeness recall some of the basics of standard cosmology. The metric of a flat homogeneous cosmology is
 $g_{\mu\nu}={\rm diag}(-1,a(t)^2,a(t)^2,a(t)^2)$, where $a(t)$ denotes the scale factor, and the classical (Friedmann) equations are,
\begin{eqnarray}
  H^2 
  = \frac{8\pi G_N}{3c^2}\rho +\frac{\Lambda}{3}
\label{Friedmann1}
\\
 \dot H = -\frac{4\pi G_N}{c^2}(p+\rho)
 \,,
\label{Friedmann2}
\end{eqnarray}
where $H(t)=\dot a/a$ denotes the Hubble parameter
($\dot a \equiv da/dt$), $\rho$ and $p$
are the energy density and pressure of the (perfect) cosmological fluid and $\Lambda$ is
the cosmological term. For simplicity we shall assume that the cosmological fluid obeys
a constant equation of state, $w=p/\rho = {\rm const.}$ and we shall take $\Lambda<0$.
From the covariant conservation equation, $\dot \rho +3H(\rho+p)=0$, we immediately infer, $\rho=\rho_0/a^{3(1+w)}$, where $\rho_0$ is an initial energy density.
Equations~(\ref{Friedmann1}--\ref{Friedmann2}) admit solutions provided
$8\pi G_N\rho_0/(3c^2) +\Lambda/3>0$, {\it i.e.} if the initial energy density of
the fluid is large enough to compensate the negative cosmological
constant~\footnote{If that condition is not met, only inhomogeneous cosmologies
are possible.}. In this case the solution of~(\ref{Friedmann1}) is
\begin{equation}
 a(t)=a_0\left[\sin\left(\sqrt{\frac{-\Lambda}{3}}\epsilon_0t\right)\right]
     ^\frac{1}{\epsilon_0}
\,;\qquad
 H(t) = \sqrt{\frac{-\Lambda}{3}}\cot\left(\sqrt{\frac{-\Lambda}{3}}\epsilon_0t\right)
\,,
\label{solution:a,H}
\end{equation}
where $\epsilon_0$ is the slow roll parameter in the case when $\Lambda=0$, {\it i.e.}: $\epsilon_0=-(\dot H/H^2)_{\Lambda\rightarrow 0}= \frac{3}{2}(1+w)$. The resulting Ricci scalar
\begin{equation}
 R(t) = 6(2H^2 + \dot H)=
 \frac{-2\Lambda}{\sin^2\left(\sqrt{\frac{-\Lambda}{3}}\epsilon_0t\right)}
    \left[2\cos^2\left(\sqrt{\frac{-\Lambda}{3}}\epsilon_0t\right)-\epsilon_0\right]
\,
\label{Ricci scalar:local}
\end{equation}
 diverges both at the Big Bang ($t_{\rm BB}=0$)  and at the Big Crunch
($t_{\rm BC}=\pi \sqrt{-3/\Lambda}\,/\epsilon_0$). Hence this Universe resembles
a closed universe. For a large $|\Lambda|$ the Universe is short lived, and thus it
does not constitute a realistic model of the Universe. The question we address next
is how to repair the model and make a realistic universe with a negative cosmological constant.

\section{Model I}
\label{Model I}

 To study the quantum backreaction in cosmology is hard, so one often resorts to studying toy models, which capture important features of the quantum backreaction in realistic cosmological settings. Our main inspiration
here are the simple model of nonlocal cosmology proposed by Deser and 
Wooodard~\cite{Deser:2007jk} (see also Refs.~\cite{Tsamis:2009ja})
and the nonlocal model considered in Ref.~\cite{Prokopec:2006yh}. We emphasize however that, apart from their simplicity, and the hope that these models incorporate some important features of
the quantum backreaction, these models are used primarily
to illustrate the main idea: to demonstrate that cosmology with the quantum backreaction included can exhibit qualitatively different behavior from standard cosmology.

To model the backreaction, we shall endow the Friedmann equation~(\ref{Friedmann1})
with the nonlocal operator $\int_{t_0}^t dt^\prime G_{\rm ret}(t;t^\prime)R(t^\prime)$,
where
\begin{equation}
 G_{\rm ret}(t;t^\prime)\equiv (\sqrt{-g}\Box )^{-1}
 =\left(\int_{t_0}^t \frac{d\tilde t}{a(\tilde t)^3}
      - \int_{t_0}^{t^\prime}\frac{d\tilde t}{a(\tilde t)^3}\right) \Theta(t-t^\prime)
\label{Green function:retarded}
\end{equation}
is the retarded Green's function of the scalar d'Alembertian operator on a homogeneous expanding background, $\Box = -(1/a^3)(d/dt)a^3(d/dt)+\nabla^2/a^2
\rightarrow  -[(d^2/dt^2) + 3H(d/dt)]$.~\footnote{Note that the Green function we use differs from the one used in~\cite{Deser:2007jk}, in the sense that our $G_{\rm ret}$ satisfies the usual property $G_{\rm ret}(t;t^\prime)=G_{\rm adv}(t^\prime;t)$, where $G_{\rm adv}$ denotes the advanced Green function, while the operator in~\cite{Deser:2007jk} does not possess this property. We have not studied the resulting
differences.} Then the nonlocal Friedmann equation~(\ref{Friedmann1}) generalizes to,
\begin{equation}
H^2 = \frac{8\pi G_N}{3c^2}\left(\frac{\rho_0}{a^{3(1+w)}}-\alpha\int_{t_0}^tdt^\prime a(t^\prime)^3R(t^\prime)\int_{t^\prime}^t\frac{d\tilde t}{a(\tilde t)^3}\right) +\frac{\Lambda}{3}
\,,
\label{Friedmann:nonlocal}
\end{equation}
where $\alpha$ is a coupling constant whose canonical dimension is $[m^{-4}]$
(because the nonlocal operator in~(\ref{Friedmann:nonlocal}) has the canonical
dimension zero).
Equation~(\ref{Friedmann:nonlocal}) can be combined with Eq.~(\ref{Friedmann2}) to yield an integral equation for $R(t)$ (an integral-differential equation for the scale factor $a(t)$), which is the equation we solve. The initial conditions are $a(t_0)=1$ and
$H(t_0)^2=(8\pi G_N \rho_0)/(3c^2)+\Lambda/3>0$. Note that the integral term
in~(\ref{Friedmann:nonlocal}) and its time derivative are both equal to zero at $t=t_0$.

 We shall now study the resulting nonlocal cosmology, which can be neatly
divided into an oscillatory epoch, an inflationary epoch, and a late time matter dominated era.

\subsection{Oscillatory Epoch}
\label{Oscillatory Epoch}

 \begin{figure}[ht]
    \begin{minipage}[t]{.8\textwidth}
        \begin{center}
\includegraphics[width=5in]{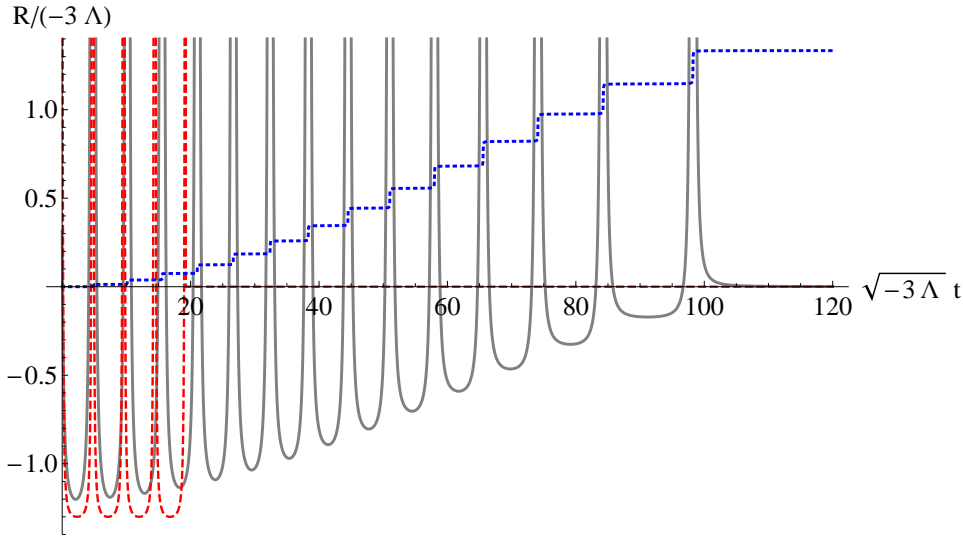}
\caption{The preinflationary (oscillatory) stage of the nonlocal cosmology~(\ref{Friedmann:nonlocal}) with a negative $\Lambda$ and matter with the equation of state parameter $w=0.3$. The solid gray line represents the Ricci scalar curvature $R(t)$, which -- as the backreaction builds up -- shows the characteristic oscillatory behavior. The backreaction term (shown in dotted blue) grows mostly during the time when the Ricci scalar is large. When the backreaction grows large enough (corresponding to a time $t\simeq 100 \sqrt{-3/\Lambda}$ for the current choice of parameters), the Universe exits from the oscillating phase end enters inflation. For comparison, we also show the first few `oscillations' of the Ricci scalar~(\ref{Ricci scalar:local}) for the corresponding local cosmology (dashed red), whose value blows up at the times $t_n=n \pi \sqrt{-3/\Lambda}/\epsilon_0$ (that signify the Big Bang and the Big Crunch singularities). For this plot we have chosen: $\alpha = 10^{-5}[-\Lambda c^2/(8\pi G_N)]$, $w=0.3$ and $\rho_0 = -{\rm e}^{3(1+w)}\Lambda c^2/(8\pi G_N)$.
\label{fig:Model1:oscillatory}}
        \end{center}
    \end{minipage}
\end{figure}
As announced above, we shall now study the evolution of an early Universe that is composed of a (large) negative cosmological term, and a matter component with a constant equation of state parameter,
 $w=p/\rho$, with $-1<w<1/3$.~\footnote{The case $w=1/3$ (radiation)
ought to be excluded because we solve the trace Friedmann equation, to which radiation
does not contribute. Likewise, a matter with $w=-1$ mimics the cosmological constant,
 and hence it also should be excluded.} To study how the nonlocal backreaction term
in~(\ref{Friedmann:nonlocal})
 modifies the Universe's evolution, it is instructive to look at the Ricci scalar
 $R(t)$, which we plot in figure~\ref{fig:Model1:oscillatory} in units of $(-3\Lambda)^{-1}$. In contrast with the
 classical local cosmology, which yields a singular $R(t)$  
 of Eq.~(\ref{Ricci scalar:local})(both at the Big Bang and Big Crunch), Eq.~(\ref{Friedmann:nonlocal}) yields completely regular evolution of $R(t)$ (solid gray line), albeit $R(t)$ reaches a large value when the local $R(t)$ from Eq.~(\ref{Ricci scalar:local}) diverges (dashed red line). In figure~\ref{fig:Model1:oscillatory} we also show the contribution of the backreaction term to the Ricci curvature (blue dotted line), from which we see that the backreaction grows mainly during the intervals when the Ricci scalar is very large (nonadiabatic regime). It is important to emphasize that the behavior shown in figure~\ref{fig:Model1:oscillatory} is generic, in the sense that the preinflationary oscillatory epoch is generic as long as initially $\Lambda<0$, $\Lambda +8\pi G_N \rho_0/c^2>0$ and $\alpha>0$. A further generic property of the model is that the oscillatory epoch ends when the backreaction grows sufficiently large such to compensate the negative cosmological term, terminating in inflation. The number and the length of oscillations depends, of course, on the choice of parameters. Clearly, for a smaller value of the backreaction coupling $\alpha$ there will be more oscillations, while the size of $|\Lambda|$ (and to a certain extent $w$) determines the length of each oscillation.
Next, we consider the inflationary epoch.

\subsection{Inflationary Epoch}
\label{Inflationary Epoch}

\begin{figure}[ht]
    \begin{minipage}[t]{.8\textwidth}
        \begin{center}
\includegraphics[width=5in]{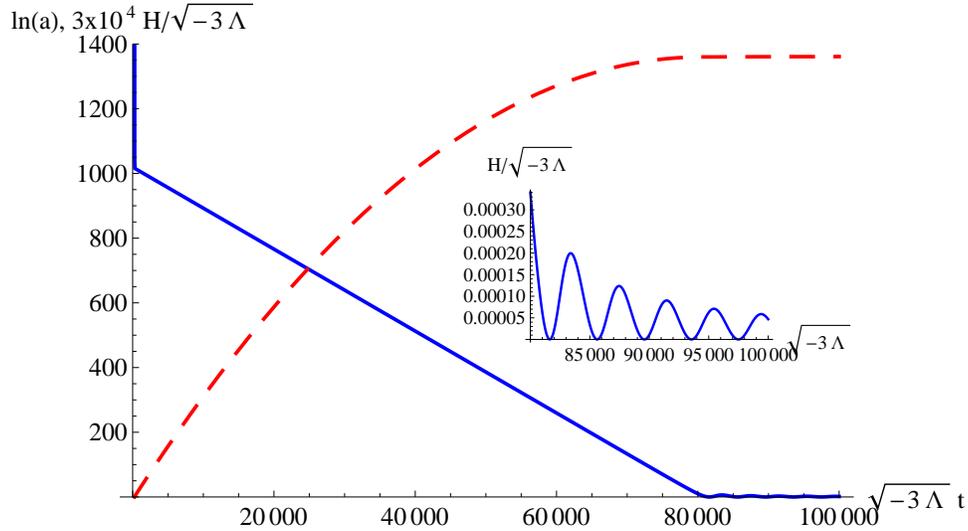}
\caption{The inflationary stage of the nonlocal cosmology~(\ref{Friedmann:nonlocal}) with a negative $\Lambda$ and a matter component with the equation of state parameter $w=0.3$. We plot $\ln(a(t))$ (red dashed line) and $H(t)$ (solid blue) as a function of time for the parameters $\alpha = 1.9\times 10^{-6}[-\Lambda c^2/(8\pi G_N)]$, $w=0.3$ and
$\rho_0 = -{\rm e}^{3(1+w)}\Lambda c^2/(8\pi G_N)$. For this choice of the parameters we get about $1260$ e-foldings of inflation. During inflation the Hubble parameter (solid blue line) decreases approximately linearly with time. After inflation ends, the Hubble parameter $H$ oscillates, as shown in the inset. During this oscillatory stage $H(t)$ remains mostly positive (corresponding to an expanding phase), although there are small intervals during which $H<0$ (corresponding to a brief contracting phase).
\label{fig:Model1:inflation}}
        \end{center}
    \end{minipage}
\end{figure}
In figure~\ref{fig:Model1:inflation} we show the evolution of the scale factor $\ln(a)$ (dashed red line) and the Hubble parameter (in units of $1/\sqrt{-3\Lambda}$) (solid blue line) during the subsequent inflationary epoch, that follows the oscillatory epoch.
There are two notable features characterizing the inflationary period in our model: (1) a linear `decay' of the Hubble parameter in cosmological time and (2) oscillations in the scale factor (and the Hubble parameter) that follow the inflationary epoch.
The linear decay of $H(t)$ could be tested by precisely measuring the dependence of
the power spectrum of scalar cosmological perturbations on momentum scale. Namely, from the slow roll approximation we know that the scalar spectrum amplitude squared (at the Hubble crossing) is given by, $H(t)^2/(8\pi^2\epsilon(t) M_P^2) = H(t)^4/(-8\pi^2 \dot H M_P^2)$, where we used $\epsilon = - \dot H/H^2$. From this formula it follows that the spectrum slope is given by $n_s-1 = -6\epsilon +2\eta \simeq  4 \dot H/H^2 = -4\epsilon$, where we made use of $\dot H={\rm  const}$~\footnote{The scalar spectrum must be generated by a dynamical adiabatic mode (still to be specified) or by an 
isocurvature mode, which could be converted after inflation into the adiabatic mode
by a curvaton-like mechanism~\cite{Lyth:2001nq}.}.
The oscillatory phase following inflation (see the inset) is reminiscent of the preinflationary oscillatory epoch. But, unlike during the preinflationary stage, here the effective cosmological constant is small and positive. These oscillations are due to the oscillations in the backreaction term and, as times goes on, they get slowly damped. As we show in the next section, after inflation the Universe enters a matter era.

The number of e-foldings in the model is quite sensitive
on the strength of the backreaction $\alpha$ in~(\ref{Friedmann:nonlocal}). In general, for smaller couplings $\alpha$ one gets longer inflation. We find that it is quite easy to get 60 (and much more) e-foldings of inflation. In the example shown in figure~\ref{fig:Model1:inflation}, $\alpha = 1.9\times 10^{-6}[-\Lambda c^2/(8\pi G_N)]$ and the number of e-foldings is about $1260$.
On the other hand, when $\alpha = 1\times 10^{-5}[-\Lambda c^2/(8\pi G_N)]$, one gets `only' $12$ e-foldings of inflation, not enough for a realistic inflationary epoch.

\subsection{Matter Era}
\label{Matter Era}

\begin{figure}[ht]
    \begin{minipage}[t]{.8\textwidth}
        \begin{center}
\includegraphics[width=5in]{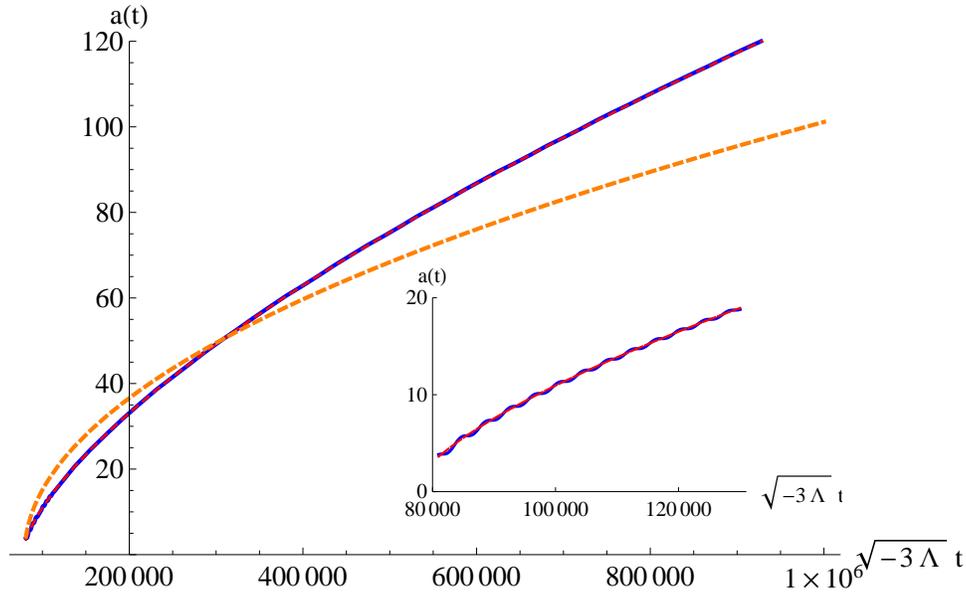}
\caption{The matter dominated stage of the nonlocal cosmology~(\ref{Friedmann:nonlocal}) with a negative $\Lambda$ and a matter component with the equation of state parameter $w=0.3$. We show the scale factor $a(t)$ (solid blue line) at late times. The scale factor follows the time dependence of a nonrelativistic matter $a\propto t^{2/3}$ (shown for illustration as the solid red line), on which small amplitude oscillations are superimposed (see the inset). The curve $a(t)\propto t^{2/3}$ can be clearly distinguished from other power law expansions. To show that, we also plot a radiation era curve (red dashed), $a(t) \propto t^{1/2}$, which clearly does not fit. The parameters are the same as in figure 2: $\alpha = 1.9\times 10^{-6}[-\Lambda c^2/(8\pi G_N)]$ and $w=0.3$.
\label{fig:Model1:matter era}}
        \end{center}
    \end{minipage}
\end{figure}
The late time evolution of the Universe in our model is characterized by a matter era, with the scale factor $a\propto t^{2/3}$, on which small oscillations are superimposed.  This can be nicely seen in figure~\ref{fig:Model1:matter era}, where the suitably adjusted matter era curve (solid red) completely overlaps with the numerical solution (solid blue curve), suggesting that the quantum backreaction term studied here can be a viable dark matter candidate. We emphasize that the nonrelativistic matter late time evolution is independent on the equation of state parameter $w$ of the preinflationary matter.
The small oscillations (that can be seen in the inset of
figure~\ref{fig:Model1:matter era}) are a memory of the inflationary stage, and characterize this model. These oscillations are slowly damped with time, such that even at very late times (in our figure $t\sim 10^6 /\sqrt{-3\Lambda}$) they are important in the sense that the Universe goes through periods of accelerated and decelerated expansion.
The question which we at this stage cannot answer is whether this feature can be used to explain the dark energy of the Universe, or perhaps to rule out the model. Whatever may be true, the postinflationary oscillations in the scale factor in a model with a realistic quantum backreaction will have to produce oscillations that are more damped than in the model studied here. This remark motivates a study of the second toy model of the quantum backreaction, which is what we do next.

\section{Model II}
\label{Model II}

\begin{figure}[ht]
    \begin{minipage}[t]{.8\textwidth}
        \begin{center}
\includegraphics[width=2.52in]{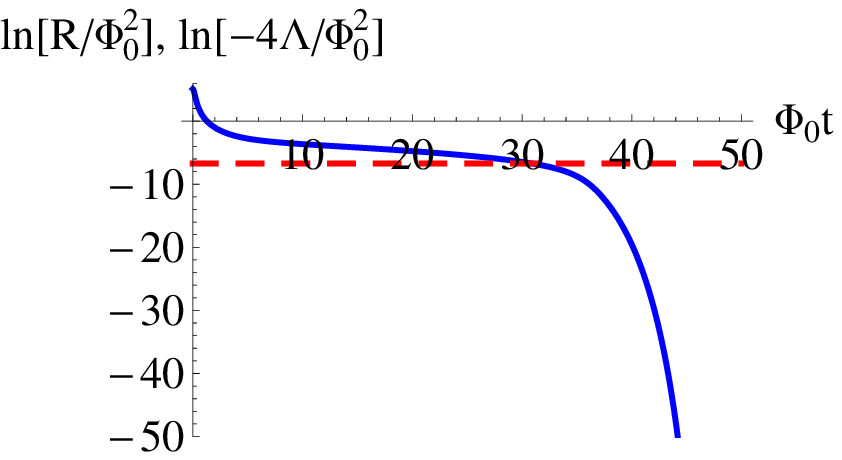}
\hskip 0.2cm
\includegraphics[width=2.52in]{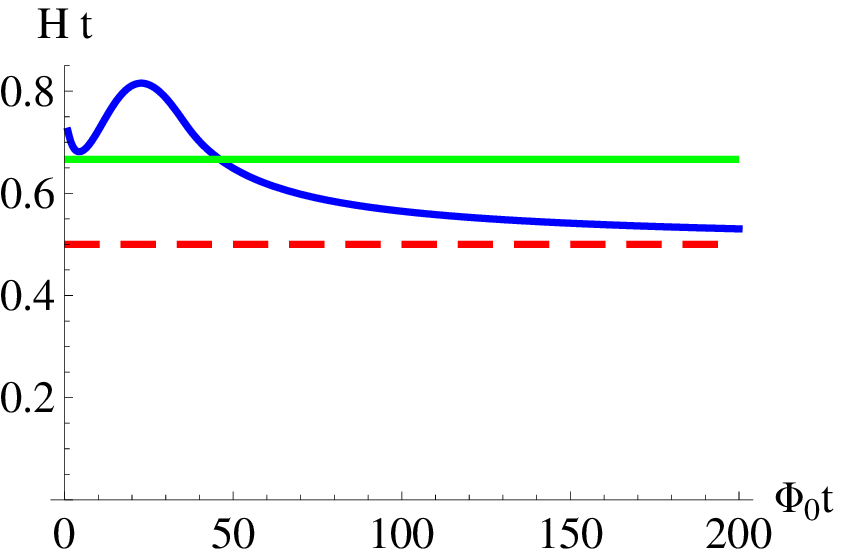}
\caption{The scalar Ricci curvature in Model II that follows
from solving Eqs.~(\ref{scalar eom}) and~(\ref{traceFriedmann:nonlocal:ModelII}).
The choice of the parameters for this plot is: $\lambda_0=0$, $\lambda_1=10^{-4}$,
$8\pi G_N\Phi_0^2/(3c^2)=10^{-1}$, $\Lambda = -10^{-3}\Phi_0^2$
 $\rho_0=50 \Phi_0^4$ and $w=0.1$, where $\Phi_0=\phi(t_0)$. From the plot on the left panel we see that $R(t)$ rapidly goes to
 zero, indicating that at late times one approaches a radiation era. This is confirmed on the right panel, where we show $tH$ (blue solid), the radiation era line $tH=1/2$ (red dashed) and the matter era line $tH=2/3$ (green solid).
\label{fig:Model2}}
        \end{center}
    \end{minipage}
\end{figure}
 Here we shall study a generalization of the nonlocal model considered in Ref.~\cite{Prokopec:2006yh}. This model has a scalar field with the potential,
\begin{equation}
V(\phi)= \frac{\lambda_0}{4!}+\frac{\lambda_1}{4!}\ln\left(\frac{\phi^4}{R^2}\right)
\,,\
\label{potential:modelII}
\end{equation}
a matter with a constant equation of state parameter,
$w= p/\rho$, and a (negative) cosmological term $\Lambda<0$, where $\lambda_1\sim \lambda_0^2$. The scalar equation of motion for this model for a homogeneous scalar then
 reads,
\begin{equation}
 \ddot \phi + 3H\dot \phi + \frac{\lambda_0}{6}\phi^3
    + \frac{\lambda_1}{6}\phi^3\left[\ln\left(\frac{\phi^4}{R^2}\right)+1\right] =0
\label{scalar eom}
\end{equation}
and the (trace) Friedmann equation for the Ricci scalar reads ({\it cf.} Eq.~(\ref{Friedmann:nonlocal})),
\begin{equation}
 R = \frac{16\pi G_N}{c^2}
 \left[-\frac{\dot\phi^2}{2}+\frac{1-3w}{2}\frac{\rho_0}{a^{3(1+w)}}
    + \frac{\lambda_0}{12}\phi^4
    + \frac{\lambda_1}{12}\phi^4\ln\left(\frac{\phi^4}{R^2}\right)
 \right]+4\Lambda
\,.
\label{traceFriedmann:nonlocal:ModelII}
\end{equation}
Solving equations (\ref{scalar eom}) and (\ref{traceFriedmann:nonlocal:ModelII}) shows
that one gets a universe which does not correspond to the singular universe with the scalar curvature given by~(\ref{Ricci scalar:local}), as would be the case in standard cosmology. Instead, $R(t)$ asymptotes to a universe with a vanishing scalar curvature, $R\rightarrow 0$ (shown on the left panel of figure~\ref{fig:Model2}), mimicking at late times a radiation era with $a\propto t^{1/2}$, as can be seen from the right panel of figure~\ref{fig:Model2}. Therefore, just like in the case of Model I, our Model~II produces a universe which, at late times, qualitatively differs from that of standard cosmology (which exhibits a Big Crunch). Instead, the quantum backreaction term in Model II compensates precisely the negative cosmological constant and the Universe asymptotes a radiation era at late times. Unlike in Model~I, Model~II neither exhibits an oscillatory nor inflationary epoch.

\section{Discussion}
\label{Discussion}

 We have considered two simple models for the quantum backreaction in cosmology. Model one consists of inserting the integral operator
 $\int^t dt^\prime G_{\rm ret}(t;t^\prime)R(t^\prime)$ into the Friedmann equation~(\ref{Friedmann:nonlocal}), while Model II makes use of the nonlocal potential $\phi^4\ln(\phi^4/R^2)$ as in Eq.~(\ref{traceFriedmann:nonlocal:ModelII}), where $G_{\rm ret}$ denotes the retarded Green function for the d'Alembertian of a homogeneous cosmology, $\phi$ a scalar field and $R$ is the Ricci scalar. We show that in both models, a universe filled initially with an arbitrary, negative cosmological term and a matter with constant equation of state parameter $w=p/\rho$, exhibits a regular late time behavior in the sense that at late times the scalar curvature remains finite and, moreover, asymptotes to zero. In Model I (Model II) the late times Ricci scalar mimics that of a matter (radiation) era. While Model I gives sensible results both for an arbitrary positive or negative cosmological constant, Model II works only for an initial negative cosmological constant.
 If either Model I or Model II could be obtained from a realistic cosmological backreaction, the mechanism presented in this letter would provide an answer to the question: why is the cosmological constant (so close to) zero. Hence, what remains to be done is to understand the structure of the quantum backreaction in a more realistic setting.

\end{document}